\documentclass[prb, twocolumn,showpacs]{revtex4}
\usepackage{graphicx}
\usepackage{amssymb}
\usepackage{amsmath}
\usepackage{color}

\begin{document}

\title{Selective excitation of plasmons superlocalized \\ at sharp perturbations of metal nanoparticles}

\author{M.V. Gorkunov$^1$, B.I. Sturman$^2$, and E.V. Podivilov$^2$}
\affiliation{
  $^1$ A.V. Shubnikov Institute of Crystallography, Russian Academy of Sciences - 119333 Moscow, Russia \\
  $^2$ Institute of Automation and Electrometry, Russian Academy of Sciences - 630090 Novosibirsk, Russia
}


\begin{abstract}Sharp metal corners and tips support plasmons localized on the scale of the
curvature radius -- superlocalized plasmons. We analyze plasmonic properties of
nanoparticles with small and sharp corner- and tip-shaped surface perturbations in terms
of hybridization of the superlocalized plasmons, which frequencies are determined by the
perturbations shape, and the ordinary plasmons localized on the whole particle. When the
frequency of a superlocalized plasmon gets close to that of the ordinary plasmon, their
strong hybridization occurs and facilitates excitation of an optical hot-spot near the
corresponding perturbation apex. The particle is then employed as a nano-antenna that
selectively couples the free-space light to the nanoscale vicinity of the apex providing
precise local light enhancement by several orders of magnitude.
\end{abstract}


\maketitle

\section{Introduction}

Efficient coupling of light to nanometer size volumes is one of the biggest challenges
of modern optics; it is desired for numerous potential applications. To overcome the
light diffraction limit, localized plasmons (LPs) supported by metal nanoparticles and
nanowires can be utilized.\cite{Novotny07,Klimov11,Schuller10} In this way, valuable
progress has been achieved in creating plasmonic optical sensors of single atoms and
molecules,\cite{Gissen11}  bio-sensors,\cite{Brolo} and nano-lasers.\cite{S1} The
related local concentration of light fields strongly enhances the Raman
scattering\cite{SERS} and the harmonic generation.\cite{SHG}

Plasmonic properties of a metal nanoparticle are primarily determined by its
shape.\cite{Novotny07,Klimov11} With analytical solutions available only for the
simplest (e.g., spherical and ellipsoidal) shapes, many efforts have been spent on
numerical modeling  of LPs of complex-shape
particles.\cite{Ruppin,Kottmann1,Kottmann2,Kelly03,Gonzalez07,Prodan, Hao,
PRB13,EPL13,JOSAB12, PRB14, Klimov2014, JOSAB14} It is recognized that adding new shape
features enriches the LP spectrum considerably. In certain cases, it is possible to
consider plasmons of complex particles as a result of hybridization of plasmons
supported by constituents of simpler shape\cite{Prodan, Hao} similarly to hybridized
diatomic electronic states.\cite{Landau3}

Metal nanoparticles and nanowires with sharp shape features, such as corners and tips,
represent an important special case. Perfectly sharp corners and tips are physically
meaningless with regard to plasmons.\cite{PRB13} The simplest polyhedral 2D particles
-- wires of smoothed square, triangular and rhomboidal cross sections --  exhibit highly
specific plasmonic properties: The resonant values of the metal permittivity are
determined by the geometric parameters of the corners -- the apex angle and the
curvature radius.\cite{EPL13,PRB13} A similar behavior was predicted for 3D particles
of smoothed cubic shapes.\cite{Klimov2014} Generally, as the sharpness increases, the
plasmons become superlocalized, i.e. localized on the scale of the curvature radius. The local values of electric field then exceed considerably those in the incident light
wave. Solid links between such superlocalized plasmons (SLPs) of metal corners and tips
and the optical singularity at perfectly sharp wedges and cones have been
established.\cite{PRB14, JOSAB14}

Nanosize perturbations of metal surface, both random\cite{Zhao2014} and
regular,\cite{JOSAB12} strongly affect plasmons and their contribution to optical
phenomena. Sharp perturbations of a flat metal surface also support SLPs which admit a relatively simple analysis including determination of
plasmonic frequencies and field distributions.\cite{JOSAB14} One can expect that a
sharp perturbation provides almost the same SLP properties whether it is placed on the flat metal surface or on a smooth nanoparticle surface. However, this expectation comes true only partially and its failure is worthy of attention.
The point is that the frequencies
of SLPs and LPs depend on essentially different shape parameters, and for this reason
they can eventually be close to each other. In this case, a strong hybridization of
large-scale LPs with short-shale SLPs of a nanoparticle can be anticipated.

In this paper, we analyze the SLP-LP hybridization and its optical consequences for the
simplest unperturbed nanoparticle shapes with non-degenerate LP spectra (elliptic wires and
spherical particles) and sharp corner- and tip-shaped surface perturbations.

\section{Basic relations}

For subwavelength particles, the full set of Maxwell's equations reduces in the
quasi-static approximation to the Laplace equation for the electric potential $\phi$
.\cite{Novotny07, Klimov11} The latter satisfies the conventional boundary conditions
including the relative (to the dielectric background) complex permittivity
$\varepsilon(\omega) = \varepsilon'(\omega) + i\varepsilon''(\omega)$. Furthermore, it
is convenient to transfer from the differential eigenvalue problem for $\phi$ in space
to an equivalent integral problem for the surface charge density
$\sigma$.\cite{Klimov11,PhilMag89,Mayergoyz05} In this approach, the plasmonic
eigenfunctions $\sigma_j$ obey the integral equation
\begin{equation}\label{Equation}
\int_B  K({\bf r},{\bf r}')\,\sigma_j({\bf r}') d{\bf r}'= \Lambda_j\, \sigma_j({\bf
r}),
\end{equation}
where the eigenvalues $\Lambda_j = (\varepsilon_j + 1)/(\varepsilon_j - 1)$ are
expressed by the real resonant permittivity $\varepsilon_j$, the kernels in the 3D case
(nanoparticle) and the 2D case (nanowire) are
\begin{equation}\label{3DKernel}
K_{\rm 3D}({\bf r},{\bf r}') = \frac{{\bf n} \cdot ({\bf r} - {\bf r}')}{2\pi \,|{\bf r}
- {\bf r}'|^3}, \ K_{\rm 2D}({\bf r},{\bf r}') = \frac{{\bf n} \cdot ({\bf r} - {\bf r}')}{\pi \,|{\bf r}
- {\bf r}'|^2},
\end{equation}
$B$ is the 3D or 2D boundary of the particle, and ${\bf n} = {\bf n}({\bf r})$ is the
unit vector of the external normal.

Although the integral operator in Eq.~\eqref{Equation} is non-symmetric, its eigenvalues
are real and its eigenfunctions are mutually orthogonal for the scalar product defined
as~\cite{Klimov11,PhilMag89}
\begin{equation}\label{scalar3D}
(\sigma_1\cdot\sigma_2)_{\rm 3D} = \int_B \ \frac{\sigma_1({\bf r})\sigma_2({\bf
r}')}{|{\bf r}-{\bf r}'|} d{\bf r}'d{\bf r},
\end{equation}
\begin{equation}\label{scalar2D}
(\sigma_1\cdot\sigma_2)_{\rm 2D} = -2\int_B \sigma_1({\bf r})\sigma_2({\bf r}')\ln(|{\bf
r}-{\bf r}'|)d{\bf r}'d{\bf r}.
\end{equation}
This allows us to normalize $\sigma_j$ as $(\sigma_j\cdot\sigma_k)= \int_B \sigma_j
\phi_kd{\bf r}' \equiv \langle\sigma_j \phi_k\rangle = \delta_{jk}$, where $\phi_k({\bf
r})$ is the surface value of the electric potential induced by the charge density
$\sigma_k$ and the brackets denote the surface integral. Note that $\phi_k$ can be also
obtained directly as a solution (denoted as $\tau_k$ in Ref.~\onlinecite{Mayergoyz05}) of the
integral problems adjoint to Eq.~\eqref{Equation}, i.e., with the arguments of the
kernels permuted.

Considering the particle in an external electric field ${\bf E}_0(\bf r)$, one can find
the coefficients of  expansion of $\sigma$ in terms of $\sigma_j$ and calculate the key
observables.\cite{Klimov11, Mayergoyz05, PRB13} In particular, evaluation of the
induced dipole moment of the particle $\langle{\bf r}\sigma\rangle$ yields the
polarizability tensor
\begin{equation}\label{polarizability}
\alpha_{\mu\nu}=\sum\limits_j\frac{[\varepsilon(\omega) - 1](\varepsilon_j - 1)}{4\pi[\varepsilon_j - \varepsilon(\omega)]}
{\langle r_\mu\sigma_j\rangle \langle \phi_j n_\nu \rangle},
\end{equation}
where $r_\mu$ and $n_\nu$ are components of $\bf r$ and $\bf n$. One can also calculate
the normal component of the total electric field above the particle surface and
determine its ratio to the external field applied in the same direction:
\begin{equation}\label{xi}
\xi({\bf r})=1+ \sum\limits_j\frac{\varepsilon_j[\varepsilon(\omega) - 1]}{\varepsilon_j - \varepsilon(\omega)}
\langle \phi_j({\bf r}')\ {\bf n}({\bf r}')\cdot{\bf n}({\bf r}) \rangle\ \sigma_j({\bf r}).
\end{equation}
The absolute value $|\xi|({\bf r})$ gives the local light-field enhancement factor.

Note the presence of the complex-valued factors $[\varepsilon_j - \varepsilon(\omega)]$
in the denominators of Eqs.~\eqref{polarizability} and \eqref{xi} that reach their
minimum absolute values $|\varepsilon''(\omega_j)|$ at the resonant frequencies
$\omega_j$ such that $\varepsilon'(\omega_j) = \varepsilon_j$. Accordingly, a strong
resonant increase of the particle response can be expected if $\varepsilon''(\omega_j)
\ll |\varepsilon'(\omega_j)|$. Often it is sufficient to restrict ourselves to a single
resonant term in Eqs.~\eqref{polarizability} and \eqref{xi}.

Numerical solution of the eigenvalue problem \eqref{Equation} is the most demanding
stage of calculations. In the 2D case, it poses a one-dimensional integral eigenvalue
problem which can be solved on a desktop computer. In the 3D case, it can be
considerably simplified by choosing appropriate particular cases, e.g., by considering a
particle of axially symmetric shape that allows surface parametrization in the spherical
coordinates as a single-valued function $r = r(\theta)$. Then one can introduce the
charge density $\sigma(\theta,\varphi)$ and search for the eigenfunctions in the form
$e^{im\varphi}\sigma_j(\theta)$ with integer $m$. For the external electric field applied
along the symmetry axis, only the modes with $m = 0$  are excited, and
Eq.~\eqref{Equation} leads to a 1D integral eigenvalue problem.

\section{Numerical results}
\subsection{Sharp corner on wire}

\begin{figure}
	\centering
	\includegraphics[width=7.5cm]{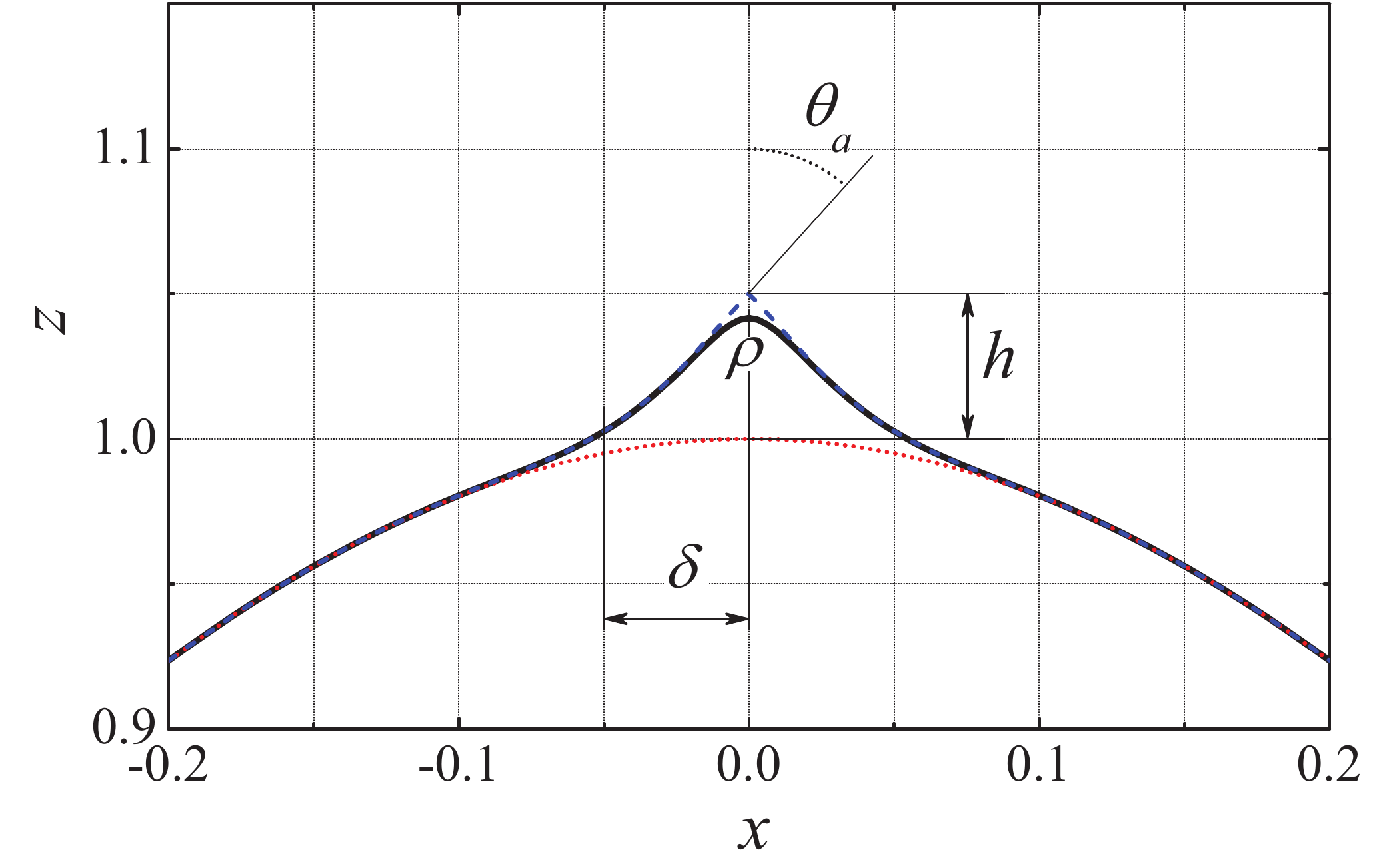}
	\caption{Profile of the corner-shaped perturbation of elliptical wire as described by Eqs.~\eqref{r0} and \eqref{r1} for $a=2$, $\delta=0.05$, $h=0.05$, and $\rho=0.01$ (solid line). The dotted and dashed lines show the unperturbed wire and ideally sharp $\rho \rightarrow 0$ perturbation profiles respectively.}
	\label{Fig1}
\end{figure}

Let us consider a wire of elliptical cross-section with a sharp corner-shaped
perturbation. The ellipticity removes the frequency degeneracy of LPs of a circular wire. The scale invariance allows for an arbitrary size normalization,
and we parameterize the unperturbed wire boundary line in the polar coordinates as
\begin{equation}\label{r0}
r_0(\varphi)=(\cos^2\varphi+a^2\sin^2\varphi)^{-1/2},
\end{equation}
where $a$ is the ellipse axis ratio. The perturbed boundary line $r(\varphi) =
r_0(\varphi) + r_1(\varphi)$ includes a small sharp corner-shaped perturbation given by
\begin{equation}\label{r1}
r_1(\varphi) = \frac{h\delta\
e^{-\varphi^2/\delta^2}}{\sqrt{\delta^2+\varphi^2}+\sqrt{p^2+\varphi^2}},
\end{equation}
where the height $h$ and the half-width $\delta$ are small positive parameters, the half
apex angle of the corner is $\theta_a = \arctan(\delta/h)$, and the sharpness is
controlled by the small parameter $p$ that determines the curvature radius $\rho =
p\delta/h$,  see Fig.~\ref{Fig1}.

\begin{figure}
\centerline{\includegraphics[width=8cm]{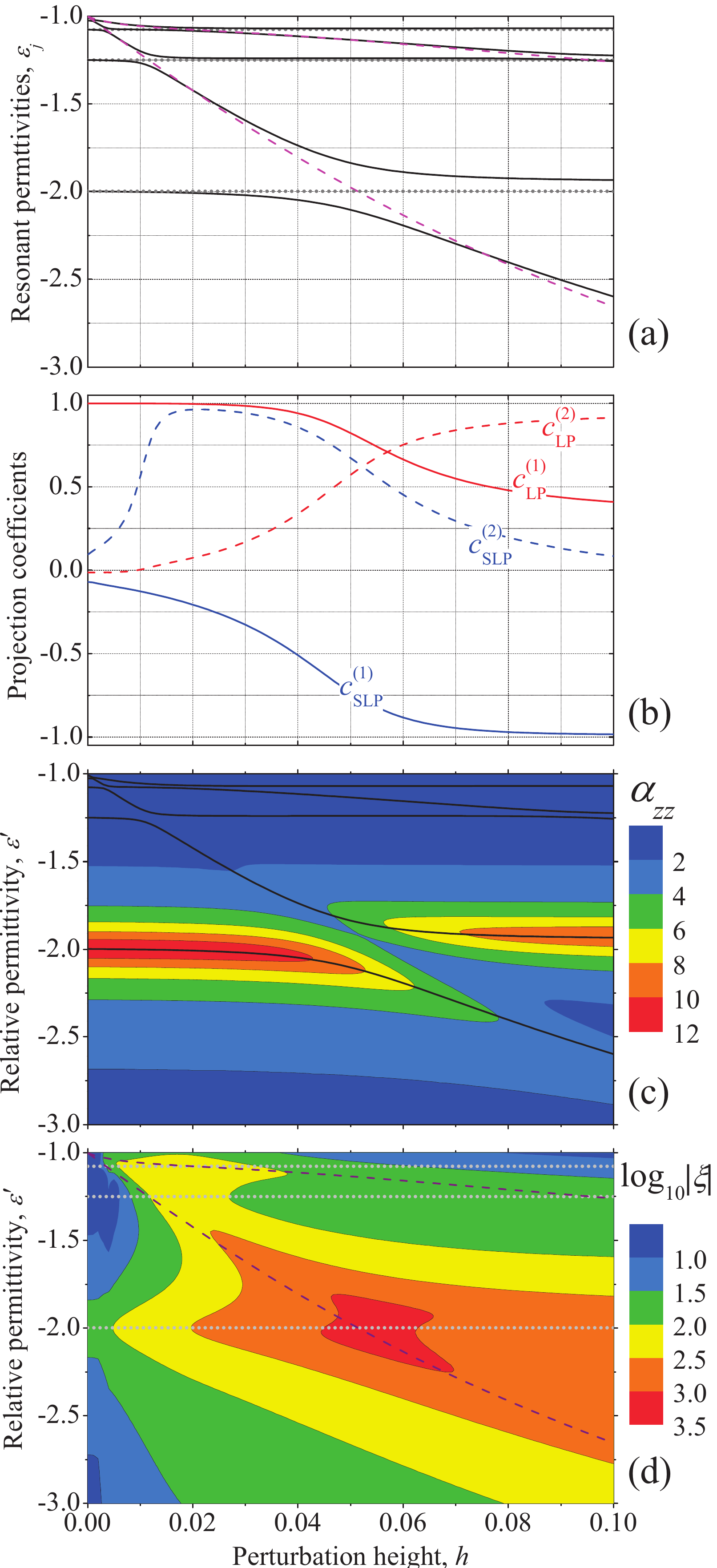}}
\caption{Plasmonic properties of elliptic wire (the axis ratio $a=2$) with the corner-shaped perturbation of the half-width $\delta=0.05$, the curvature radius $\rho=0.01$ and variable height $h$. (a): The lowest resonant permittivity branches (solid) compared to those for the same perturbation of plane surface \cite{JOSAB14} (dashed) and those of the unperturbed wire (dotted). (b): Projection coefficients for the plasmons with the lowest (solid) and second-lowest (dashed) resonant permittivities.(c): Color map of the wire polarizability with the resonant permittivity branches shown by lines. (d): Color map of the field enhancement factor with the SLP- and LP-branches shown by dashed and dotted lines respectively. The ratio $|\varepsilon'|/\varepsilon''=20$ was assumed for (c) and (d).
}\label{abc2D}
\end{figure}

With this parametrization we solved numerically Eq.~\eqref{Equation} using a
discretization with crowding of points near the corner apex. The typical number of
points about $1000$ was sufficient, and it was made sure that the general modal
properties -- zero total charge, twin symmetry,\cite{Mayergoyz05} and orthogonality --
were accurately fulfilled.

Four lowest branches $\varepsilon_j(h)$ for the perturbed elliptic cross-section are
shown in Fig.~\ref{abc2D}a by solid lines. The intersecting horizontal dotted lines and negatively
tilted dashed lines, given for comparison, show a few lowest resonant permittivity branches for
the unperturbed wire (LP-branches) and for the same sharp perturbation placed on flat
surface\cite{JOSAB14} (SLP-branches), respectively. Each $\varepsilon_j(h)$ branch stays close either to an LP-branch or to an SLP-branch except for the vicinities of their intersections. Evidently, in these regions we have a strong hybridization of localized and
superlocalized plasmons with the avoided crossing structure of the
branches typical for hybridized states.\cite{Landau3}

Furthermore, we have found that the surface-charge distributions $\sigma_j(\varphi)$ are
close to superpositions of $\sigma_{\rm LP}(\varphi)$ and $\sigma_{\rm SLP}(\varphi)$
corresponding to the LP resonances of the unperturbed wire and the SLP resonances of the
perturbation placed on a flat surface:
\begin{equation}\label{mixing}
\sigma_j\approx c^{(j)}_{\rm LP}\sigma_{\rm LP} + c^{(j)}_{\rm SLP}\sigma_{\rm SLP}.
\end{equation}
The mixing coefficients can be  evaluated as projections $c^{(j)}_{\rm
LP}=(\sigma_j\cdot\sigma_{\rm LP})$ and $c^{(j)}_{\rm SLP}=(\sigma_j\cdot\sigma_{\rm
SLP})$. As seen from Fig.~\ref{abc2D}b, the surface charge distributions are very close
either to $\sigma_{\rm LP}$ (when $|c^{(j)}_{\rm LP}|\approx 1$ and $|c^{(j)}_{\rm
SLP}|\ll 1$) or to $\sigma_{\rm SLP}$ (when $|c^{(j)}_{\rm SLP}|\approx 1$ and
$|c^{(j)}_{\rm LP}|\ll 1$) when the $\varepsilon_j$ branches are close to LP- and SLP-branches respectively.

Near the avoided crossings, the LP and SLP contributions to $\sigma_j$ are comparable.
Here the wire optical properties -- the polarizability and the field enhancement factor
-- behave very peculiarly. A representative color map of the polarizability in
Fig.~\ref{abc2D}c shows that a noticeable dipole response of the wire occurring only
when the relative permittivity is close to that of the lowest LP resonance is suppressed
by the perturbation near the avoided crossing. At the same time, the field enhancement
factor $|\xi|$ experiences there  a dramatic increase by several orders of magnitude
(see Fig.~\ref{abc2D}d).

\begin{figure}
\centerline{\includegraphics[width=8cm]{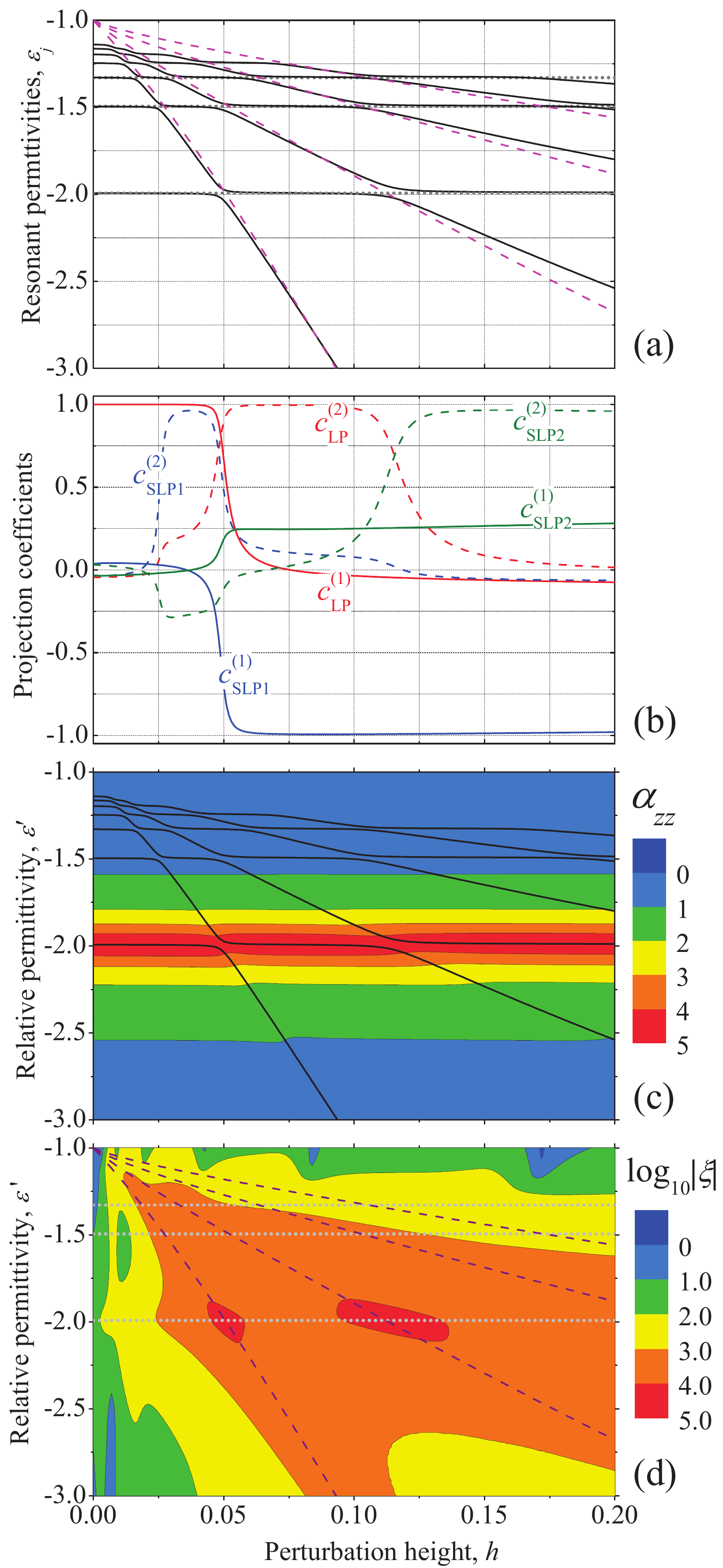}}
\caption{Plasmonic properties of the spherical particle with the tip-shaped perturbation of the half-width $\delta=0.1$, the  curvature radius $\rho=0.01$ and variable height $h$. (a): The lowest resonant permittivity branches (solid) compared to those for the same perturbation of plane surface \cite{JOSAB14} (dashed) and those of the unperturbed sphere (dotted). (b): Projection coefficients of the plasmons with the lowest (solid) and second-lowest (dashed) resonant permittivities. (c): Color map of the polarizability with the resonant permittivity branches shown by lines. (d): Color map of the field enhancement factor with the SLP-and LP-branches shown by dashed and dotted lines respectively. The ratio $|\varepsilon'|/\varepsilon''=20$ was assumed for (c) and (d).
}\label{abc3D}
\end{figure}

\subsection{Sharp tip on spherical particle}

In the 3D case, a very similar situation occurs for a small sharp tip-shaped perturbation of a smoothly shaped nanoparticle. As the simplest example we consider a spherical particle with a conical tip of the same cross-section as above, i.e., of the shape obtained by rotating the profile in Fig.~\ref{Fig1} around the vertical axis. Accordingly, the particle surface is parameterized in the spherical coordinates by the polar angle dependent radius vector $r(\theta)=r_0(\theta)+r_1(\theta)$ where the functions $r_0$ and $r_1$ are given by Eqs.~\eqref{r0} and \eqref{r1} and $a=1$.

The obtained branches of the resonant permittivity shown in Fig.~\ref{abc3D}a follow either the horizontal LP-branches of the unperturbed sphere or the tilted SLP-branches of the conical tip on a flat surface. Near the intersections of the LP- and SLP-branches, the avoided crossings indicating strong hybridization occur. We evaluated the projection coefficients $c^{(j)}_{\rm LP}$ and $c^{(j)}_{\rm SLP 1,2}$ using the charge distribution $\sigma_{\rm LP}$ of the lowest (dipolar) LP resonance of the unperturbed sphere and distributions $\sigma_{\rm SLP 1}$ and $\sigma_{\rm SLP 2}$ of the lowest (SLP1) and second-lowest (SLP2) resonances of the tip-shaped perturbation of a flat surface respectively. As illustrated by Fig.~\ref{abc3D}b, the plasmon charge densities are very close either to $\sigma_{\rm LP}$ or to $\sigma_{\rm SLP 1,2}$ and a noticeable LP-SLP mixing occurs near the avoided crossings.

\begin{figure}
\centerline{\includegraphics[width=8cm]{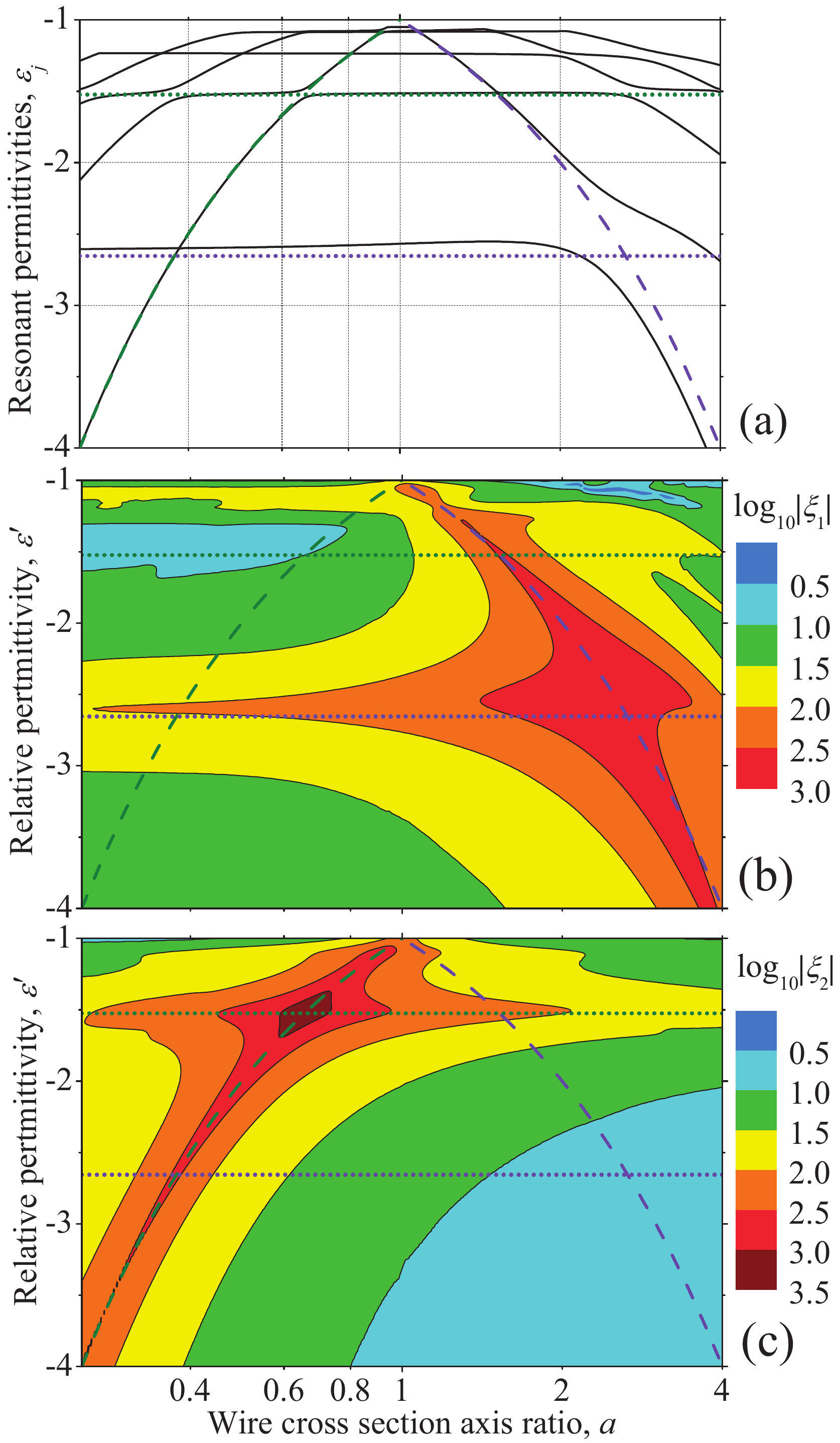}}
\caption{Plasmonic properties of elliptic wire of variable cross section axis ratio $a$ with a pair of corner-shaped perturbations of the same half-width $\delta=0.05$ and curvature radius $\rho=0.01$ and different heights, $h_1=0.1$ and $h_2=0.025$, located at $\varphi=0$ and $\varphi=\pi/2$ respectively. (a): The lowest resonant permittivity branches (solid) compared to those for the SLPs of the same perturbations of plane surface \cite{JOSAB14} (horizontal dotted) and for the LPs of elliptical wire (dashed). (b) and (c): Color maps of the local field enhancement factor on perturbations 1 and 2 respectively; the lowest SLP-branches for the same perturbations on plane surface and the LP-branches of the wire are shown as in (a);  $|\varepsilon'|/\varepsilon''=20$ is assumed.}\label{TwoRidges}
\end{figure}

The colormaps of polarizability and field enhancement factor presented in Figs.~\ref{abc3D}c and \ref{abc3D}d respectively also show qualitative similarity with the 2D-case. Quantitatively,
the band of high polarizability here is less affected by the perturbation showing a smaller decrease near the avoided crossings. The difference stems apparently from the weaker LP-SLP interaction in the 3D case (for the same perturbation size) as the tip-shaped perturbation is localized in all directions. Note that the field enhancement factor here is much higher (for the same perturbation curvature radius) and reaches the values of the order of $10^5$ near the avoided crossings.

\subsection{Two sharp corners on wire}

To clarify the situation when two sharp perturbations are present, we consider a wire of elliptical cross section  with a well separated pair of corner-shaped perturbations of the form \eqref{r1}. We assume that both perturbations have equal widths and curvature radii but different heights, $h_1$ and $h_2$, and are located at $\varphi=0$ and $\varphi=\pi/2$ respectively. To demonstrate the sensitivity of the SLP excitation to the unperturbed wire shape, we consider the variable axis ratio $a$.

As seen in Fig.~\ref{TwoRidges}a, the resonant permittivity branches here also form avoided crossings near the intersections of the wire LP-branches (the lowest are given by $\varepsilon=a^{\pm 1}$) and the SLP-branches of the perturbations (independent of the wire cross section and represented by the horizontal lines). The local field enhancement at the perturbation apexes increases dramatically near the avoided crossing points as seen in Figs.~\ref{TwoRidges}b and \ref{TwoRidges}c. Importantly, the difference in the perturbation height grants a well pronounced spectral isolation of these hot-spots and no interaction between them can be traced from our data.

\section{Discussion}

As we have seen, each small sharp perturbation of the nanoparticle shape supports plasmons superlocalized near the perturbation apex. Their frequencies
are determined by the corresponding perturbation shape. When one of SLP frequencies is close to
that of the ordinary LPs (localized on the whole particle) a strong LP-SLP hybridization
occurs enabling an LP-SLP synergy. The latter is very beneficial for the light-particle interactions: 
Large resonant excitation cross-section of LPs is combined with the high
local-field concentration in SLPs. As a result, the whole particle plays the role of a
nano-antenna that couples free-space light to the nanoscale volume at the top of the
perturbation.

When several small sharp perturbations are present, they are generally independent from
each other. Accordingly, in real metal nanoparticles with possibly numerous small shape
irregularities, only those SLPs will be selectively excited by light whose
frequencies are close to the whole particle LP frequencies and only the corresponding  perturbations will contribute to optical phenomena.

The conditions of selective SLP excitation are sensitive to the external conditions. By
adjusting the background permittivity by 10--20$\%$, one can vary the
local-field values by orders of magnitude (see Figs.~\ref{abc2D}d, \ref{abc3D}d, \ref{TwoRidges}b, and
\ref{TwoRidges}c). A similar effect can be achieved by changing the close environment,
e.g. by displacing the adjacent nanoobjects. This paves the way for targeted
design of multi-functional nanoscale optical systems based on metal particles with
corner- and tip-shaped features.


The chosen three-scale hierarchy of the particle shape, $\rho\ll\delta\ll1$, implying smallness of the
perturbations and their sharpness, is crucial for our considerations. It ensures the compactness of the hybridization regions (especially
in the 3D case) and huge near-field enhancement near the perturbation apexes. 
Thus, for example, downgrading to a two-scale shape with $\rho=\delta$ results in a dramatic drop of $|\xi|$ by several orders of magnitude.
Practically, the chosen scale hierarchy can be realized with the curvature radius of $1$~nm, the corner/tip
height and width in the range of $5-10$~nm, and the particle size about $100$~nm. Finally, the
assumed ratio $|\varepsilon'|/\varepsilon'' = 20$ is a good estimate for silver in the
visible, gold in the infrared, and aluminum in the
ultraviolet.\cite{JC,Palik,Rakic}

\section{Conclusion}

Plasmon resonances of metal
nanoparticles with small sharp shape perturbations 
are hybrids of plasmons of the two elementary types: the superlocalized plasmons
(SLPs) supported of the perturbations and the localized plasmons (LPs) of the unperturbed particles. An efficient optical excitation of an SLP is possible due to a strong LP-SLP hybridization when the SLP frequency is close to an LP frequency. Then the whole particle acts as a nano-antenna that selectively couples
the free-space light to a nanoscale optical hot-spot at the top of the corresponding perturbation where the light fields are enhanced by several orders of magnitude.

\acknowledgments The work was supported by the Russian Academy of Sciences via
Program 24 and by the Russian Foundation for Basic Research, project No. 13-02-12151 ofi$_-$m.


\begin{thebibliography}{0}

\bibitem{Novotny07}
L.Nolvotny and B.Hecht, {\em Principles of Nano-Optics}, Cambridge University Press
(2007).

\bibitem{Klimov11}
V. Klimov, {\em Nanoplasmonics}, Pan Stanford Publishing (2011).

\bibitem{Schuller10}
J.A. Schuller, E.S. Barnard, W. Cai, Y.C. Jun, J.S. White, and M.L.
Brongersma, Nature Mater. {\bf 9}, 193 (2010).

\bibitem{Gissen11}
N. Liu, M.L. Tang, M. Hentschel, H. Giessen and A.P. Alivisatos, Nature
Mater. {\bf 10}, 631 (2011).

\bibitem{Brolo}
A.G. Brolo, Nature Phot.{\bf 6}, 709 (2012).

\bibitem{S1}
D.J. Bergman and M.I.  Stockman, Phys. Rev. Lett. {\bf 90}, 027402 (2003).

\bibitem{SERS}
K. Kneipp, M. Moskovits, and H. Kneipp, {\em Electromagnetic Theory of SERS}, Springer (2006).

\bibitem{SHG}
A. Bouhelier, M. Beversluis, A. Hartschuh, and L. Novotny, Phys. Rev. Lett. {\bf 90}, 13903 (2003).

\bibitem{Ruppin}
R. Ruppin, Z. Phys. D {\bf 36}, 69 (1996).

\bibitem{Kottmann1}
J.P. Kottmann and O.J.F. Martin, Appl. Phys. B {\bf 73}, 299 (2001).


\bibitem{Kottmann2}
J.P. Kottmann, O.J.F. Martin, D.R. Smith, and S. Schultz, Phys. Rev. B
{\bf 64}, 235402 (2001).

\bibitem{Kelly03}
K.L. Kelly, E. Coronado, L.L. Zhao, and G.C. Schatz, J. Phys. Chem. B {\bf 107}, 668 (2003).

\bibitem{Gonzalez07}
A.L. Gonzalez and C. Noguez, J. Computational and Theoretical Nanoscience {\bf 4},
231 (2007).

\bibitem{Prodan}
E. Prodan, C. Radloff, N.J. Halas and P. Nordlander, Science {\bf 302}, 419 (2003).

\bibitem{Hao}
F. Hao, C.L. Nehl , J.H. Hafner  and P. Nordlander,
Nano Lett. {\bf 7}, 729 (2007).

\bibitem{PRB13}
B. Sturman, E. Podivilov, and M. Gorkunov,
Phys. Rev. B {\bf 87}, 115406 (2013).

\bibitem{EPL13}
B. Sturman, E. Podivilov, and M. Gorkunov,
EPL {\bf 101}, 57009 (2013).

\bibitem{Klimov2014}
V. Klimov, G.-Y. Guo, and M. Pikhota, J. Phys. Chem. C {\bf 118}, 13052 (2013).

\bibitem{PRB14}
B. Sturman, E. Podivilov, and M. Gorkunov,
Phys. Rev. B {\bf 89}, 045429 (2014).

\bibitem{JOSAB14}
B. Sturman, E. Podivilov, and M. Gorkunov,
J. Opt. Soc. Am. B {\bf 31}, 1607 (2014).

\bibitem{JOSAB12}
B. Sturman, E. Podivilov, and M. Gorkunov,
J. Opt. Soc. Am. B {\bf 29}, 3248 (2012).

\bibitem{Landau3}
L.D. Landau and E.M. Lifshits, {\em Quantum mechanics}, Pergamon Press, Oxford
(1991).

\bibitem{Zhao2014}
Y. Zhao, X. Liu, D.Y. Lei , and Y. Chai
Nanoscale {\bf 6}, 1311 (2014).

\bibitem{PhilMag89}
F. Ouyang and M. Isaacson
Phil. Mag. B {\bf 60}, 481492 (1989).

\bibitem{Mayergoyz05}
I.D. Mayergoyz, D.R. Fredkin, and Z. Zhang, Phys. Rev. B {\bf 72}, 155412
(2005).

\bibitem{JC}
P.B. Johnson and R.W. Christy,
Phys. Rev. B {\bf 6}, 4370 (1972).

\bibitem{Palik}
D.W. Lynch  and W.R. Hunter, in {\em Handbook of Optical Constants of
Solids},  ed. by E.D. Palik, {Academic Press, New York} (1985).

\bibitem{Rakic}
A.D. Rakic, A.B. Djurisic, J.M. Elazar, and M.L. Majewski, Appl. Opt.
{\bf 37}, 5271 (1998).

\end{thebibliography}
\end{document}